\documentclass[preprint]{aastex}
\usepackage[usenames,dvipsnames]{color}
\usepackage{natbib,graphicx,latexsym,lscape,pdflscape,url}
\usepackage[breaklinks,colorlinks,urlcolor=blue,citecolor=black,linkcolor=blue]{hyperref} 

\newcommand{\Spitzer}{{\sl Spitzer}}

\newcommand{\WISE}{{\sl WISE}}

\newcommand{\Lsun}{\mbox{$L_{\sun}$}}

\newcommand{\Mjup}{\mbox{$M_{\rm Jup}$}}
\newcommand{\Rjup}{\mbox{$R_{\rm Jup}$}}
\newcommand{\degree}{\mbox{$^{\circ}$}}
\newcommand{\perpix}{\mbox{pixel$^{-1}$}}

\newcommand{\masyr}{\hbox{mas\,yr$^{-1}$}}

\newcommand{\Mtot}{\mbox{$M_{\rm tot}$}}

\newcommand{\Lbol}{\mbox{$L_{\rm bol}$}}

\newcommand{\mbol}{\mbox{$m_{\rm bol}$}}
\newcommand{\Mbol}{\mbox{$M_{\rm bol}$}}
\newcommand{\Teff}{\mbox{$T_{\rm eff}$}}
\newcommand{\logg}{\mbox{$\log(g)$}}

\newcommand{\obj}{WISE~J0146+4234}
\newcommand{\objA}{WISE~J0146+4234A}
\newcommand{\objB}{WISE~J0146+4234B}
\newcommand{\objAB}{WISE~J0146+4234AB}
\newcommand{\objlong}{WISE~J014656.66+423410.0}
\newcommand{\dpc}{\hbox{$10.6^{+1.3}_{-1.8}$\,pc}}

\newcommand{\zoelong}{WISEPC~J140518.40+553421.4}
\newcommand{\joelong}{UGPS~J072227.51$-$054031.2}
\newcommand{\guslong}{WISEPC~J014807.25$-$720258.7}
\newcommand{\kaylong}{WISEPC~J205628.90+145953.3}
\newcommand{\acelong}{WISEPC~J121756.91+162640.2}
\newcommand{\janlong}{WISEPA~J041022.71+150248.5}
\newcommand{\elilong}{CFBDSIR~J145829+101343}
\newcommand{\boblong}{WISEPA~J173835.53+273258.9}
\newcommand{\maelong}{WISEPA~J154151.66$-$225025.2}
\newcommand{\donlong}{ULAS~J003402.77$-$005206.7}
\newcommand{\redlong}{WISE~J073444.02$-$715744.0}
\newcommand{\flolong}{WISE~J033515.01+431045.1}
\newcommand{\suelong}{WISEPA~J213456.73$-$713743.6}
\newcommand{\jeblong}{WISEPC~J232519.54$-$410534.9}
\newcommand{\vallong}{WISE~J030449.03$-$270508.3}

\newcommand{\zoe}{WISE~J1405+5534}
\newcommand{\joe}{UGPS~J0722$-$0540}	      
\newcommand{\gus}{WISE~J0148$-$7202}   
\newcommand{\kay}{WISE~J2056+1459}     
\newcommand{\ace}{WISE~J1217+1626}    
     
\newcommand{\eli}{CFBDSIR~J1458+1013}	      
\newcommand{\bob}{WISE~J1738+2732}     
   
\newcommand{\don}{ULAS~J0034$-$0052}

\newcommand{\sue}{WISE~J2134$-$7137}   
\newcommand{\jeb}{WISE~J2325$-$4105}

\shorttitle{Low-Luminosity Binary at the T/Y Transition}
\shortauthors{Dupuy et al.}

\begin{document}

\title{Discovery of a Low-Luminosity, Tight Substellar Binary at the T/Y Transition\altaffilmark{*,\dag}}

\author{Trent J.\ Dupuy,\altaffilmark{1}
        Michael C.\ Liu,\altaffilmark{2} and
        S.\ K.\ Leggett\altaffilmark{3}}

      \altaffiltext{*}{Data presented herein were obtained at the
        W.M.\ Keck Observatory from telescope time allocated to the
        National Aeronautics and Space Administration through the
        agency's scientific partnership with the California Institute
        of Technology and the University of California. The
        Observatory was made possible by the generous financial
        support of the W.M.\ Keck Foundation.}

      \altaffiltext{\dag}{Some observations were obtained under
        program ID GN-2012B-DD-5 at the Gemini Observatory, which is
        operated by the Association of Universities for Research in
        Astronomy, Inc., under a cooperative agreement with the NSF on
        behalf of the Gemini partnership: the National Science
        Foundation (United States), the National Research Council
        (Canada), CONICYT (Chile), the Australian Research Council
        (Australia), Minist\'{e}rio da Ci\^{e}ncia, Tecnologia e
        Inova\c{c}\~{a}o (Brazil) and Ministerio de Ciencia,
        Tecnolog\'{i}a e Innovaci\'{o}n Productiva (Argentina).}

      \altaffiltext{1}{The University of Texas at Austin, Department
        of Astronomy, 2515 Speedway C1400, Austin, TX 78712, USA}

      \altaffiltext{2}{Institute for Astronomy, University of Hawai`i,
        2680 Woodlawn Drive, Honolulu, HI 96822 USA}

      \altaffiltext{3}{Gemini Observatory, Northern Operations Center,
        670 N.\ A'ohoku Place, Hilo, HI 96720 USA}

\begin{abstract}

  We have discovered that the brown dwarf \objlong\ is a close binary
  ($0\farcs0875\pm0\farcs0021$, $0.93^{+0.12}_{-0.16}$\,AU) from Keck
  laser guide star adaptive optics imaging. Our photometry for this
  system reveals that both components are less luminous than those in
  any known substellar binary. Combining a new integrated-light
  spectrum (T9p) and resolved $YJH$-band photometry from Keck allows
  us to perform spectral decomposition and assign component types of
  T9 and Y0. Many of the unusual features in the spectrum might be
  explained by high surface gravity: $Y$-band peak broadened to the
  blue; $J$-band peak broadened to the red; $H$-band peak shifted
  slightly to the red; and red $Y-J$ colors. Interestingly, the very
  low component luminosities imply that the T9 primary is unexpectedly
  cold ($\Teff = 345\pm45$\,K assuming an age of 10\,Gyr), making it
  $\approx$100\,K cooler than any other late-T dwarf and comparable to
  Y~dwarfs.  One intriguing explanation for this apparent discrepancy
  is that the $J$- and $H$-band spectral features that trigger the
  transition from T to Y spectral types are highly gravity-dependent.
  This can be tested directly in the very near future by orbit
  monitoring.  We constrain the orbital period to be $\lesssim$10\,yr
  by combining evolutionary model-based mass estimates for the
  components ($\approx$12--21\,\Mjup, 1$\sigma$ at 10\,Gyr) with a
  statistical constraint on the semimajor axis ($\lesssim$1.3\,AU).
  Such a period is shorter than any other known T/Y transition binary,
  meaning that \objAB\ will likely yield a dynamical mass within the
  next few years.

\end{abstract}


\section{Introduction}

The last few years have seen a rapid increase in the discovery of
brown dwarfs significantly cooler than any previously known objects in
the solar neighborhood \citep{2011ApJ...730L...9L,
  2011ApJ...740..108L, 2011ApJS..197...19K, 2012ApJ...753..156K,
  2014ApJ...786L..18L}. \citet{2011ApJ...743...50C} presented the
discovery of several brown dwarfs from \textsl{Wide-field Infrared
  Survey Explorer} (\WISE) data, and they distinguished these objects
from warmer brown dwarfs primarily using spectral features in the $J$
and $H$ bands and proposed a new ``Y'' spectral class for the sample.
\citet{2013Sci...341.1492D} presented the first comprehensive sample
of parallaxes for the Y~dwarfs and derived bolometric luminosities
showing that even objects with quite different near-infrared spectra
(and thus very different \Teff\ estimates from model atmospheres) in
fact had indistinguishable temperatures of $\approx$400--450\,K.
Assuming ages of 1--5\,Gyr, typical for field brown dwarfs,
evolutionary models predict masses of 6--20\,\Mjup\ for the
Y~dwarfs. Thus, Y~dwarfs could be free-floating planetary-mass
($\lesssim$13\,\Mjup) objects, but this can only be tested directly
with dynamical mass measurements from binaries. However, none of the
binaries yet discovered at the T/Y transition
\citep{2011AJ....142...57G, 2011ApJ...740..108L, 2012ApJ...758...57L}
have short enough estimated periods ($\lesssim$20\,yr) to enable orbit
determinations in the near future.

We present here the discovery of a tight ($0\farcs09$) binary at the
Y~dwarf boundary from our ongoing Keck laser guide star adaptive
optics (LGS AO) imaging survey. \citet{2013ApJS..205....6M} discovered
the unresolved source \objlong, hereinafter \obj, and assigned a
spectral type of Y0 based on Keck/NIRSPEC spectroscopy in $J$ and $H$
bands (also see \citealp{2012ApJ...753..156K}).  Using a new, higher
signal-to-noise ratio (S/N) Gemini/GNIRS spectrum covering
0.9--2.4\,\micron\ we derive an integrated-light spectral type of T9p
for \obj.  Adopting the parallax distance of \dpc\ from
\citet{2014ApJ...783...68B}, the projected separation of \objAB\ is
$\approx$0.9\,AU, which would make it tighter than any previously
known L or T~dwarf visual binary at discovery.

\section{Observations \label{sec:keck}}

\subsection{Keck/NIRC2 LGS AO \label{sec:keck}}

We first observed \obj\ during a period of exceptionally good seeing
on 2012~Sep~7~UT with the facility near-infrared camera NIRC2 with the
LGS AO system at the Keck~II telescope \citep{2006PASP..118..297W,
  2006PASP..118..310V}.  According to the differential image motion
monitor (DIMM) at the Canada-France-Hawaii Telescope, the seeing in
the visible was $0\farcs35$ on Mauna Kea during our observations.
Compared to all archived seeing measurements during 2012, such
favorable conditions occur $<$5\% of the time that the DIMM is active.
During our dither sequences we kept the LGS centered in NIRC2's narrow
camera field of view.  The wavefront sensor recorded an LGS flux
equivalent to a $V \approx 9.8$--10.2~\,mag star and a lower bandwidth
sensor monitored the star 2MASS~J01465144+4234388 ($R = 15.3$\,mag,
65\arcsec\ away from the target) in order to perform tip-tilt
corrections.  We later observed \obj\ in the same setup on 2012~Oct~8
and 2013~Oct~22~UT, during times of unusually good seeing
($\approx0\farcs30$--$0\farcs40$ and $0\farcs45$--$0\farcs55$,
respectively).

\begin{figure}

\begin{flushleft}
\centerline{
\includegraphics[height=1.6in,angle=0]{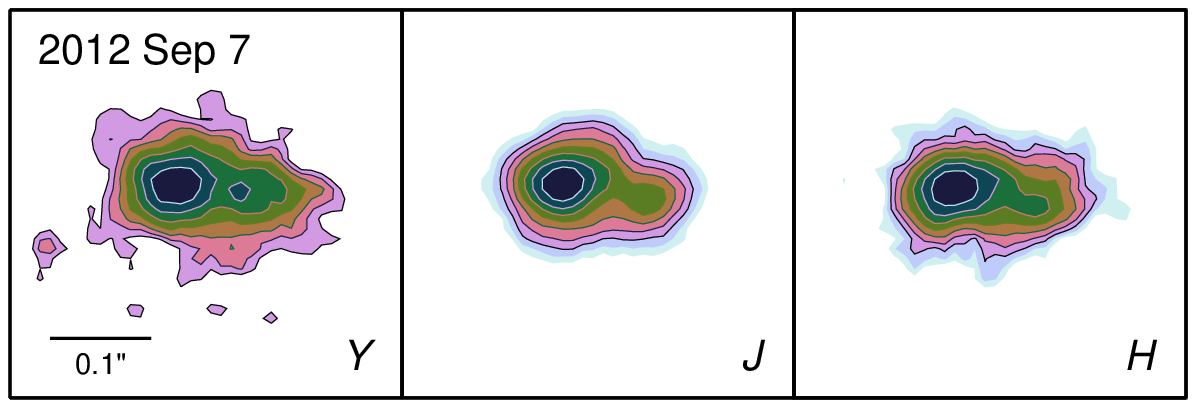}
\includegraphics[height=1.6in,angle=0]{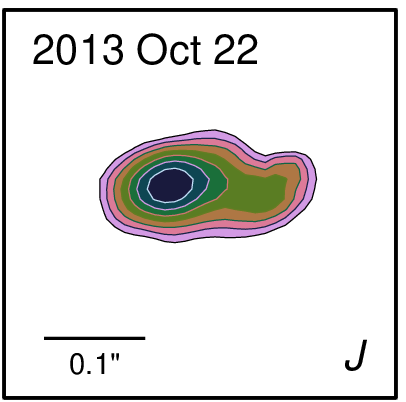}}
\vskip 0.05in
\centerline{
\includegraphics[height=1.6in,angle=0]{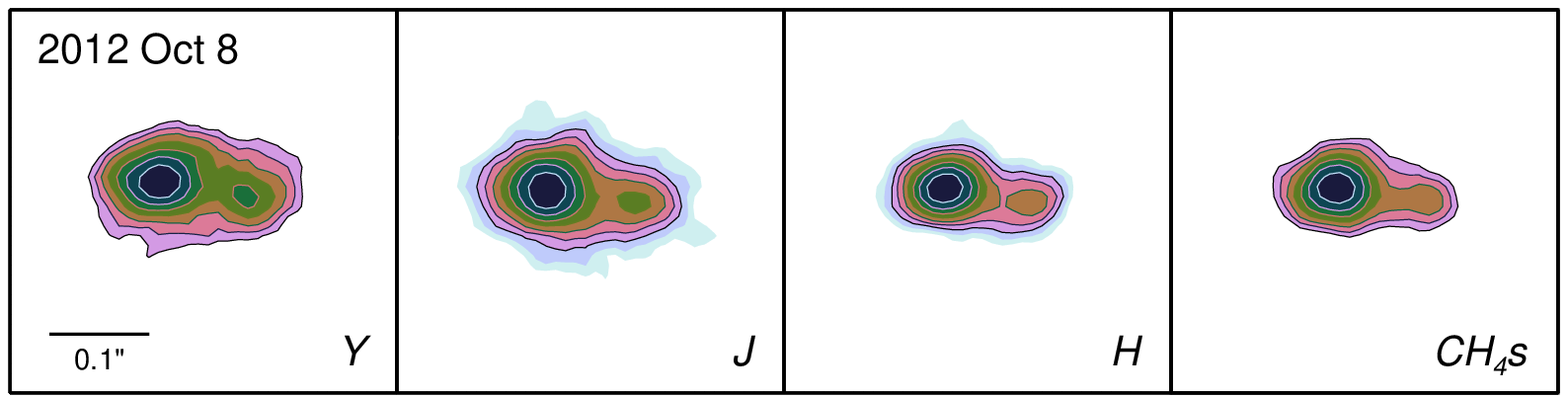}}
\end{flushleft}

\caption{\normalsize Contour plots of our Keck LGS AO images from
  which we derive astrometry and flux ratios (Table~\ref{tbl:keck}).
  Contours are in logarithmic intervals from unity to 10\% of the peak
  flux in each band.  Images have been rotated such that north is
  up. \label{fig:keck}}

\end{figure}

We reduced our LGS AO data in the same fashion as our previous work
\citep{2008ApJ...689..436L, 2009ApJ...692..729D, 2009ApJ...699..168D,
  2010ApJ...721.1725D}.  Reduced images from each data set are shown
in Figure~\ref{fig:keck}.  We measured binary parameters by fitting
three-component Gaussians to individual images and determined the
uncertainties in these parameters from the rms scatter among each data
set.  To correct for non-linear distortions in NIRC2 we used the
calibration of \citet{2010ApJ...725..331Y}, from which we also adopt
linear terms of $9.952\pm0.002$\,mas\,\perpix\ for the pixel scale and
$+0\fdg252\pm0\fdg009$ for the detector's $+y$-axis orientation
relative to North.  Table~\ref{tbl:keck} presents the binary
separation, position angle (P.A.), and flux ratio that we derived from
each data set, along with the airmass and AO-corrected FWHM of our
images.  We note that the astrometry measured in different bandpasses
at a given epoch agree well within the errors and have reasonable
$\chi^2$, supporting our adopted astrometric uncertainties.

\subsection{Gemini/NIRI Photometry \label{sec:niri}}

\obj\ had no previously published photometry on the Mauna Kea
Observatories (MKO) system, and the 2MASS system photometry from
\citet{2012ApJ...753..156K} has very large errors, so we obtained new
MKO photometry using the facility camera NIRI at the Gemini-North
Telescope \citep{2003PASP..115.1388H} on 2013~Jan~10 and
2013~Jan~12~UT.  All data were taken on photometric nights, with
seeing typically around 0$\farcs$8 and UKIRT FS~6 used for photometric
calibration.  Our observations at $Y$, $J$, $H$, $CH_4s$, and $K$
bands are summarized in Table~\ref{tbl:gem}.  We note that our
photometry is highly inconsistent with the values presented by
\citet{2012ApJ...753..156K}, even after accounting for the conversion
from 2MASS to MKO systems. $J_{\rm MKO}-J_{\rm 2MASS} = -0.260$\,mag
and $H_{\rm MKO}-H_{\rm 2MASS} = 0.074$\,mag according to synthesized
photometry from our GNIRS spectrum (Section~\ref{sec:gnirs}).  Our
photometry is 1.6\,mag (6.0$\sigma$) fainter at $J_{\rm MKO}$ band and
2.5\,mag (9.3$\sigma$) fainter at $H_{\rm MKO}$ band.  The cause of
this discrepancy is not clear, but we note that
\citet{2013ApJ...763..130L} reported similar cases for three Y~dwarfs
that they found were significantly fainter (0.5--1.0\,mag) than
published values from \citet{2011ApJ...743...50C} and
\citet{2012ApJ...753..156K}.  We suspect large systematic errors in
the photometry from \citet{2012ApJ...753..156K} for \obj\ because
colors based on those values were noted by \citet{2013A&A...550L...2L}
to be extremely unusual ($z-J$ and $z-H$ colors $\gtrsim$2\,mag redder
than objects of similar spectral type), whereas our $J$ and $H$
magnitudes resolve this discrepancy.  \citet{2014ApJ...783...68B} have
also reported $H$-band photometry ($H_{\rm MKO} = 20.91\pm0.21$\,mag)
that agrees with our higher precision Gemini measurement.  We
therefore use our photometry in the following analysis.

\subsection{Gemini/GNIRS Spectroscopy \label{sec:gnirs}}

\begin{figure}

\centerline{
\includegraphics[width=6.5in,angle=0]{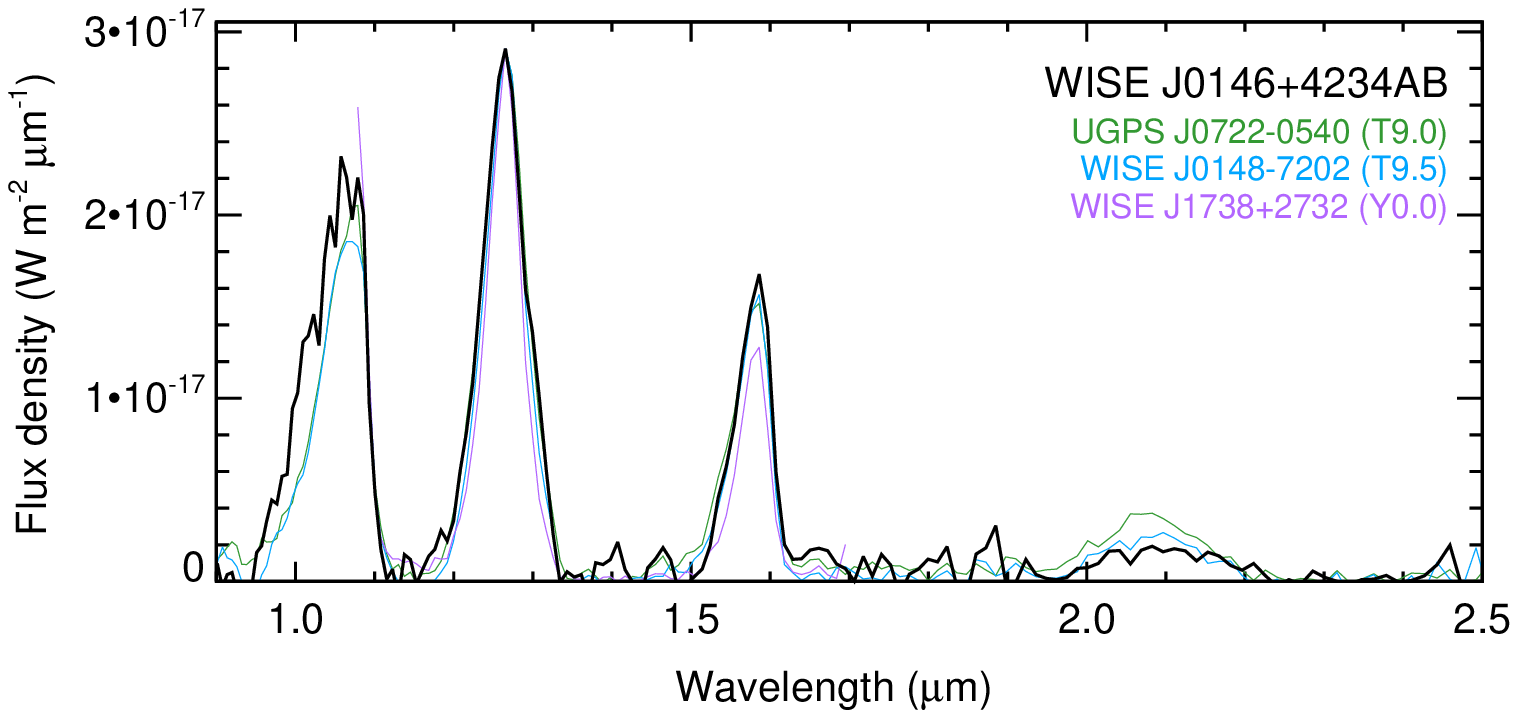}}
\centerline{
\includegraphics[width=2.3in,angle=0]{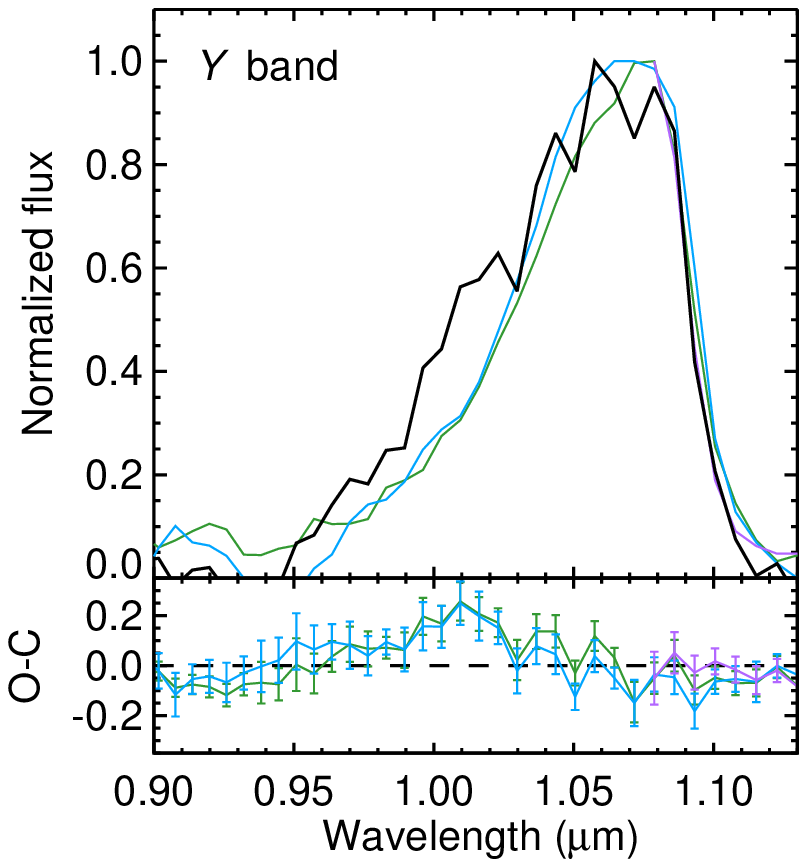}
\hskip -0.3in
\includegraphics[width=2.3in,angle=0]{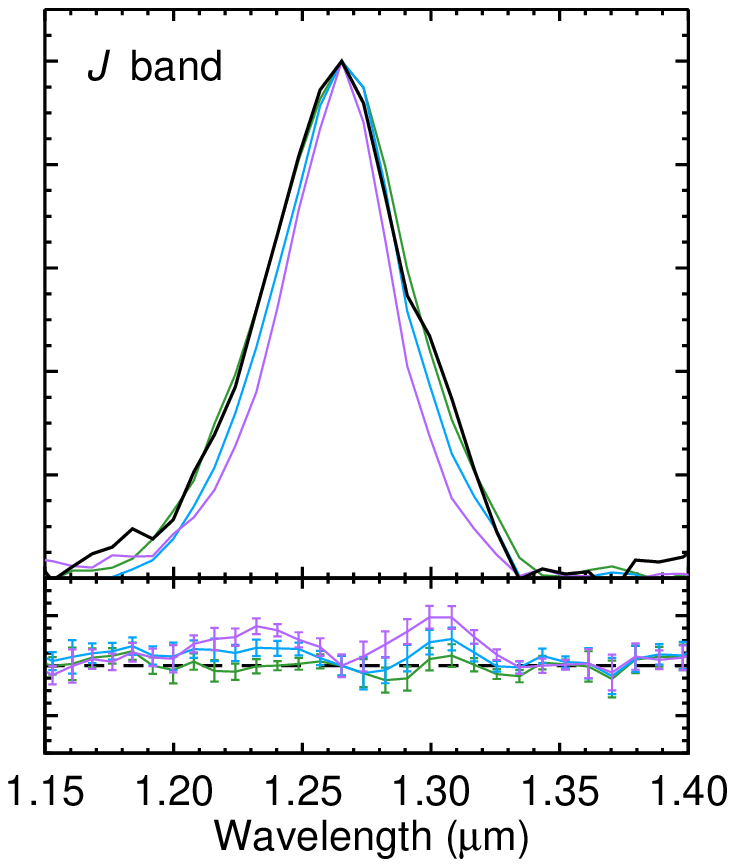}
\hskip -0.3in
\includegraphics[width=2.3in,angle=0]{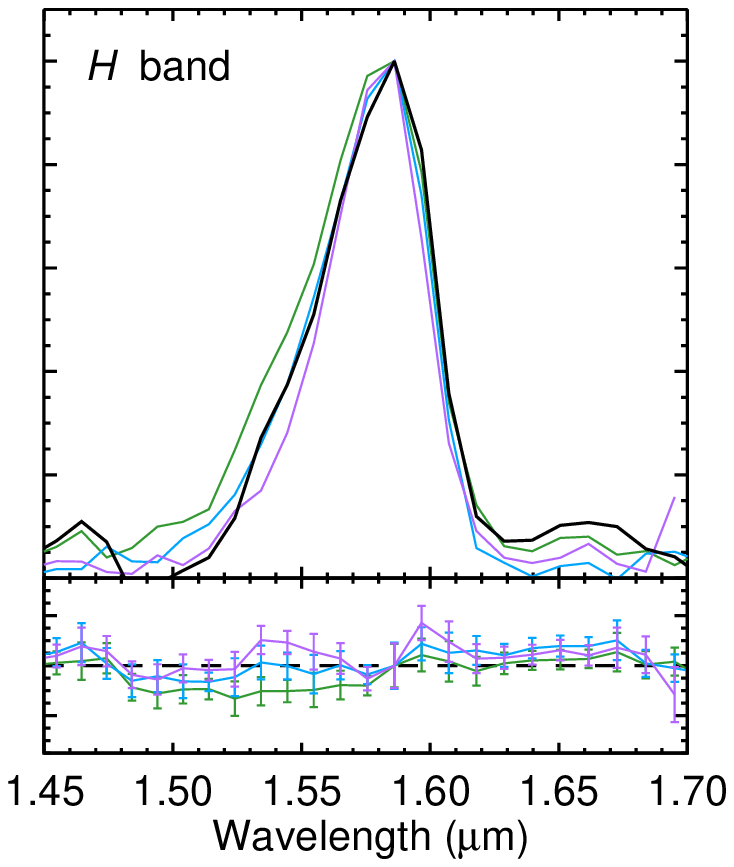}}

\caption{\normalsize \emph{Top:} Integrated-light spectrum of \objAB\
  obtained with Gemini/GNIRS (smoothed to $R\approx150$; black).  The
  spectrum is flux calibrated using our Gemini/NIRI photometry
  (Table~\ref{tbl:gem}).  Plotted for comparison are spectral
  standards that have been normalized to the peak flux of \objAB\ at
  1.15--1.40\,\micron.  \emph{Bottom:} Zoomed in plots of our spectrum
  compared to the same standards, where all spectra are instead
  normalized to their peak in that band.  Differences between our
  observed spectrum and the standards are shown with error bars in the
  bottom subpanels.  We plot one data point per resolution
  element. \label{fig:spectrum}}

\end{figure}

We obtained a full 0.9--2.4\,\micron\ spectrum of \obj\ using the
GNIRS spectrograph \citep{2006SPIE.6269E.138E} because the published
spectrum from \citet{2013ApJS..205....6M} had limited wavelength
coverage and low S/N. We used the 32 lines~mm$^{-1}$ grating,
$0\farcs15$\,\perpix\ camera, and $0\farcs675$ slit to achieve a
resolving power of $R \approx 750$ over a total integration time of
8400\,s (Table~\ref{tbl:gem}). We reduced our data in a similar
fashion to our previous work \citep[e.g.,][]{2014ApJ...780...62L}, and
the flat fielded, sky subtracted, and rectified integrated-light
spectrum of \objAB\ is shown in Figure~\ref{fig:spectrum}. We flux
calibrated the spectrum using our NIRI photometry at $Y$, $J$, and $H$
bands (Table~\ref{tbl:gem}). To account for modest discrepancies
between our synthesized and measured $YJH$ colors, we added 0.2\,mag
errors in quadrature to our synthesized photometry to achieve
$p(\chi^2) = 0.5$ when computing the optimal scaling factor and its
error.

From our calibrated spectrum we computed $K$-band photometry as check
against our very low S/N NIRI photometry ($K = 22.4\pm0.4$\,mag).
After accounting for the uncertainty in the calibration, we
synthesized $K = 21.31\pm0.24$\,mag. This is inconsistent with our
NIRI photometry at 2.3$\sigma$. Both our photometry and the $K$-band
portion of our GNIRS spectrum are very low S/N, and our synthesized
photometry may be subject to systematic errors associated with
extrapolating our flux calibration from $YJH$ to $K$. Therefore, in
the following analysis we adopt an average value of $K =
21.75\pm0.25$\,mag that is consistent at 1.3$\sigma$ with both
measured and synthesized photometry.

\section{Results}

\subsection{Companionship \label{sec:comp}}

From our resolved photometry alone, companionship is highly likely.
The blue near-infrared colors of the companion \objB\ are very similar
to \objA, implying that it not likely to be a background giant. The
difference in $CH_4s-H$ color of $0.29\pm0.15$\,mag between the two
components implies that they have similar levels of methane absorption
and therefore are both late-type T or Y dwarfs. According to
\citet{2012ApJ...758...57L}, the probability of finding such an
unassociated late-T dwarf in the entire $10\farcs2\times10\farcs2$
field of view of NIRC2's narrow camera is $1.6\times10^{-6}$, thus the
likelihood that the two components of \objAB\ are not physically bound
is negligible. Moreover, the proper motion of
$(\mu_{\alpha}\cos{\delta}, \mu_{\delta}) = (-0\farcs441 \pm
0\farcs013, -0\farcs026 \pm 0\farcs016)$\,yr$^{-1}$ and parallax of
$0\farcs094\pm0\farcs014$ measured by \citet{2014ApJ...783...68B}
implies that an unassociated background star would move by
$0\farcs56\pm0\farcs02$ between our first and last observations, which
is strongly ruled out by our astrometry (Table~\ref{tbl:keck}).

\subsection{Spectral Types \label{sec:spt}}

To determine the integrated-light spectral type of \objAB\ we directly
compared our spectrum to standards from \citet{2011ApJ...743...50C}.
For reference, we also computed commonly used spectral indices, which
we report in Table~\ref{tbl:indices}. In determining spectral types we
focus on the $J$- and $H$-band regions of the spectrum, leaving the
somewhat unusual $Y$-band region to our later discussion in
Section~\ref{sec:logg}.

The relative heights of the flux peaks in the $J$ and $H$ bands better
match the T9 standard \joelong\ and T9.5 dwarf \guslong\ than the Y0
standard \boblong\ (Figure~\ref{fig:spectrum}).  The $J$-band region,
normalized to its peak flux, best matches the T9 standard \joe\
because the Y0 standard, and even the T9.5 dwarf \gus, have notably
narrower flux peaks. This is reflected by the fact that $J$-band
spectral indices W$_J$ \citep{2007MNRAS.381.1400W} and $J$-narrow
\citep{2013ApJS..205....6M} best match the T9 dwarfs from
\citet{2013ApJS..205....6M}. However, the peak-normalized $H$-band
region seems to be intermediate between T9 and Y0 standards, best
matching the T9.5 dwarf \gus.  In terms of indices, this manifests as
an NH$_3-H$ index \citep{2008A&A...482..961D} that best matches T9.5
dwarfs. We also note that the wavelength at the peak of the $H$-band
flux is shifted slightly to the red by 0.006\,\micron, a feature that
has only previously been observed in \zoelong, which was typed as
Y0(pec?)  by \citet{2011ApJ...743...50C}. No two of these three traits
(T9-like $J$ band, T9.5-like $H$ band, and shifted $H$-band peak) have
been observed in any other object, making the integrated-light
spectrum of \objAB\ particularly unusual. The two methane indices
\citep[defined by][]{2006ApJ...637.1067B} are broadly consistent with
a late-T or Y spectral type.  The CH$_4-J$ index best matches T9.5
dwarfs but is also consistent with Y0 dwarfs.  The CH$_4-H$ index is
problematic because the numerator has almost no flux in it by the
late-T dwarfs.  Thus, even though our spectrum has sufficient S/N to
accurately measure this index, many comparison objects do not, so our
value ends up being consistent within the rms of all T8.5--Y0 spectral
type bins.

Overall, we find evidence for both T9 and T9.5 spectral types by
visually comparing to spectral standards and examining spectral
indices. Most published work has focused on $J$ band in determining
spectral types at the T/Y transition, and our $J$ band spectrum leans
more toward a type of T9. Therefore, we assign a type of T9p to the
integrated-light spectrum of \objAB. In this case, the peculiar
designation refers to both the 0.006\,\micron\ shift in the $H$-band
peak, as originally suggested by \citet{2011ApJ...743...50C} for \zoe,
and the T9.5-like $H$-band spectrum.

To determine spectral types for the individual components of \objAB,
we performed spectral decomposition following the method outlined in
Section~5.2 of \citet{2012ApJS..201...19D}. Briefly, we consider all
possible pairs from a library of template spectra and find the optimal
scale factors needed for a pair of templates to best match the
observed integrated-light spectrum. We then compute synthetic relative
photometry for these pairings and compute the $\chi^2$ of these values
compared to the measured flux ratios from our Keck LGS AO imaging.
Unlike in \citet{2012ApJS..201...19D}, here we have a much smaller
library of template spectra given the number of objects later than T8
with high quality near-IR spectra. Our library comprises four T8.5
dwarfs (\donlong, \acelong{A}, ULAS~J133553.45+113005.2, and
Wolf~940B), two T9 dwarfs (\joe\ and \elilong{AB}), one T9.5 dwarf
(\gus), five Y0 dwarfs (\janlong, \ace{B}, \zoe, \bob, and \kaylong),
and one Y0.5 \\ (\maelong). Spectra for these objects were published
by \citet{2007MNRAS.381.1400W}, \citet{2008MNRAS.391..320B,
  2009MNRAS.395.1237B}, \citet{2011ApJ...743...50C}, and
\citet{2014ApJ...780...62L}.  Another difference in our analysis
compared to \citet{2012ApJS..201...19D} is the wavelength range
tested. Because many published spectra do not extend into $Y$ or $K$
band, we use only the 1.15--1.85\,\micron\ region for our fitting.

\begin{figure}

\centerline{\includegraphics[width=6.0in,angle=0]{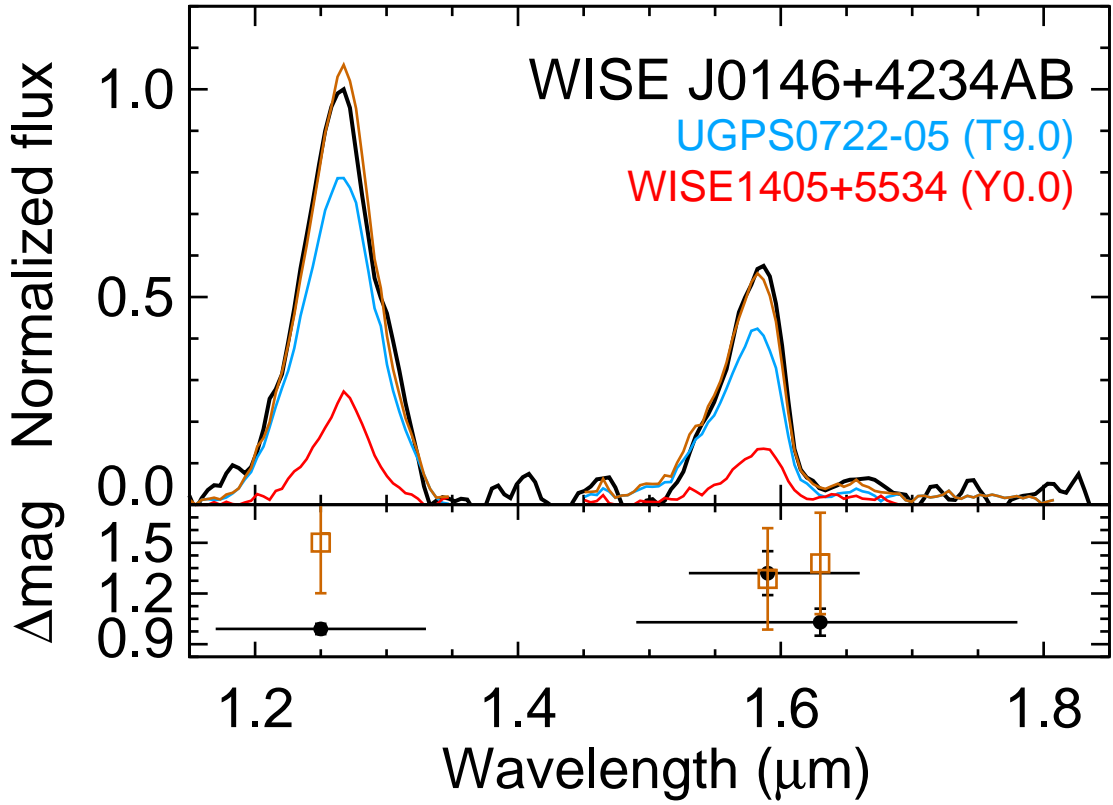}}

\caption{\normalsize Integrated-light spectrum (black) and best
  matching component templates (colored lines).  The bottom subpanel
  shows the observed $J$-, $H$-, and $CH_4s$-band broadband flux
  ratios used to constrain the decomposition (filled black circles
  with errors) and the resulting flux ratios computed from the best
  matching template pair (open brown squares). \label{fig:decomp}}

\end{figure}

The best matching template pair, both in terms of fitting our
integrated-light spectrum and flux ratios at $J$, $H$, and $CH_4s$
bands, is \joe\ (T9) and \zoe\ (Y0p). A primary spectral type of T9 is
consistent with the fact that this brighter component should dominate
the integrated-light spectrum given our Keck flux ratios of
$\approx$1.0\,mag in $J$ and $H$ bands. It is not surprising that
\zoe\ provides the best matching secondary component because of some
of the features it shares with our integrated-light spectrum (shifted
$H$-band peak and large CH$_4-H$ index). Figure~\ref{fig:decomp} shows
this best-fit template pairing.  A number of other template pairings
give both a good fit to the spectrum and reasonable flux ratios. These
either use \joe\ (T9) or a T8.5 (\don\ or \ace{A}) as the primary and
a Y0 (\ace{B}, \bob, or \kay) or the T9.5 \gus\ as the
secondary. Therefore, we assign component types corresponding to our
best-fit template pair (T9+Y0) with uncertainties of $\pm$0.5 subtypes
in each.  Given some of the unusual features seen in integrated light,
we note that one or both components may also be peculiar despite our
not typing them as such.

\subsection{Near- and Mid-Infrared Colors and Magnitudes \label{sec:color}}

\begin{figure}

\centerline{
\includegraphics[width=3.2in,angle=0]{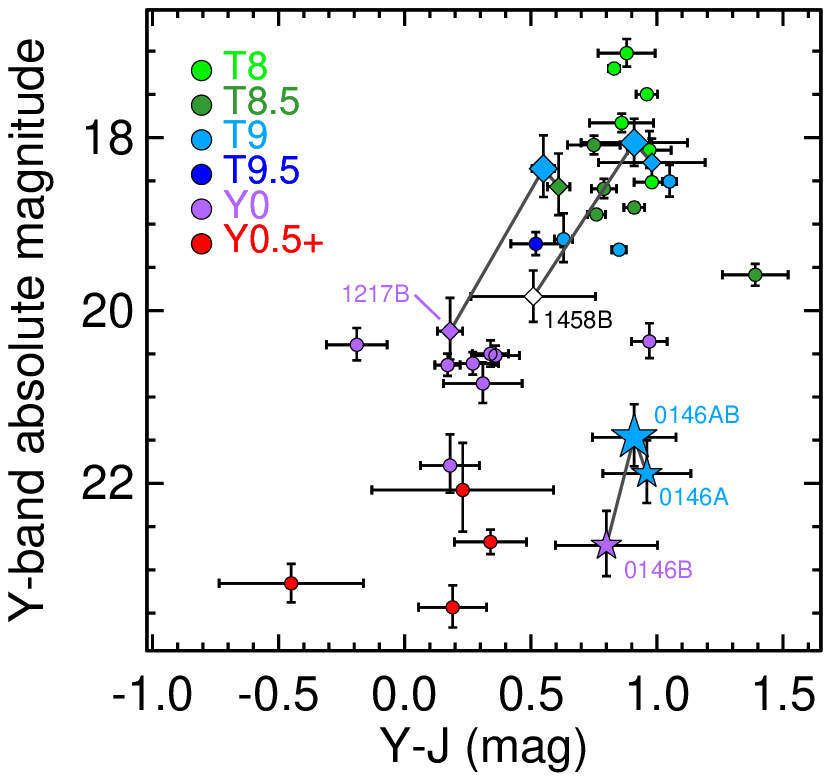}
\includegraphics[width=3.2in,angle=0]{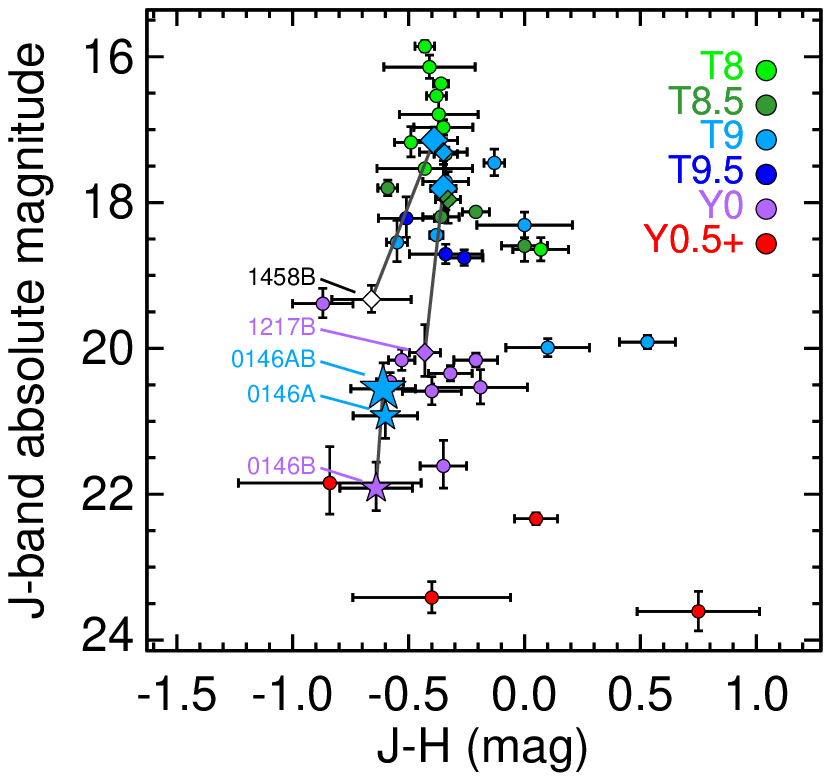}}
\centerline{
\includegraphics[width=3.2in,angle=0]{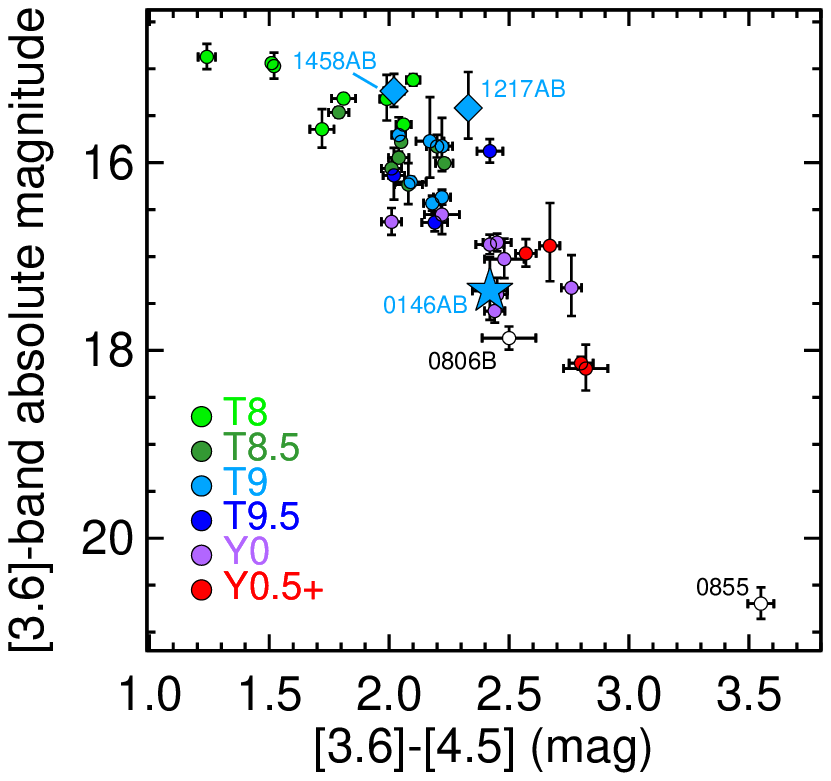}
\includegraphics[width=3.2in,angle=0]{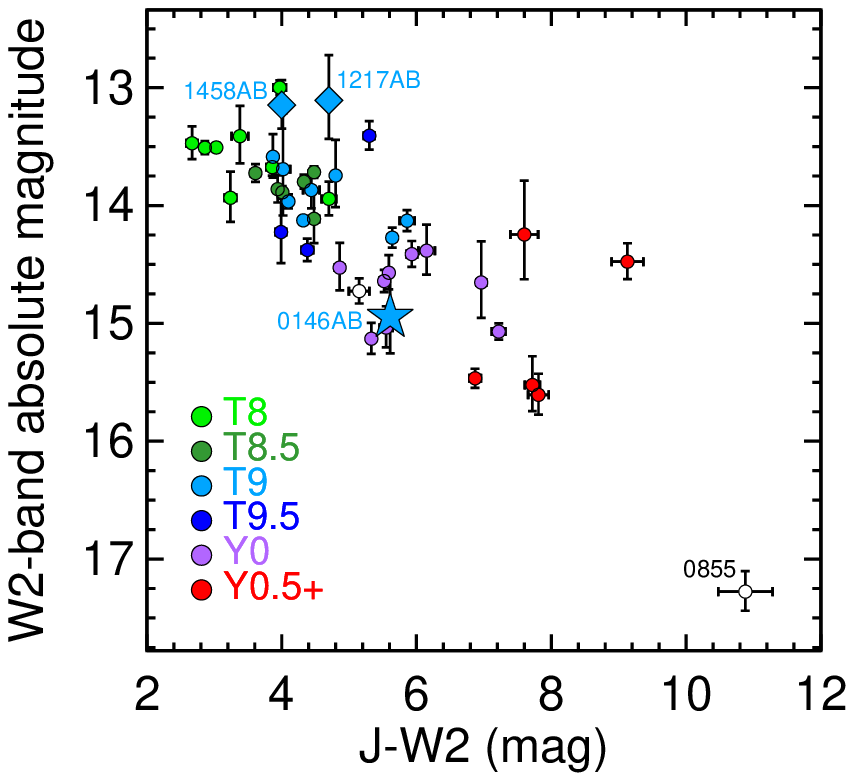}}

\caption{\normalsize Color--magnitude diagrams for all objects with
  spectral types T8 and later that have accurate photometry
  \citep[][and references therein]{2014arXiv1411.2020L} and direct
  distance measurements \citep{2014ApJ...783...68B,
    2013Sci...341.1492D, 2012ApJS..201...19D, 2012ApJ...748...74L,
    2011ApJ...730L...9L, 2014ApJ...796....6L, 2014ApJ...796...39T,
    2013AJ....145...84W}.  Binary systems are shown in both integrated
  light and as resolved components when possible.  \objAB\ and its
  components are plotted as star symbols, showing unusual absolute
  magnitudes (and sometimes colors) for their spectral types.  Other
  binaries are plotted as diamonds.  All data points are color coded
  according to spectral type, with open/white points indicating that
  no spectra are available.  Near-infrared photometry is on the Mauna
  Kea Observatory (MKO) system, and objects with distance modulus
  errors $>$0.5\,mag are not plotted for clarity.  \label{fig:cmd}}

\end{figure}

Our new photometry and spectral analysis allow us to compare the
colors of \objAB\ to other late-T and Y~dwarfs (Figure~\ref{fig:cmd}).
The $YJH$ colors of both components and the integrated-light $J-K$
color are mostly typical of other T8.5--Y0 objects.  The color in
which the components appear most unusual is $Y-J$, where the Y0
secondary's color is $0.80\pm0.20$\,mag.  This would not be unusual
for a late-T dwarf, but it is on the extreme red end of colors for
Y~dwarfs where only one comparably red object is known, the Y0 dwarf
\redlong\ \citep[$Y-J = 0.97\pm0.07$\,mag;][]{2014arXiv1411.2020L}.
Moreover, both components of \objAB\ are distinctive in that they are
the reddest known objects in $Y-J$ at faint absolute magnitude ($M_Y
\gtrsim 21$\,mag).  Their $J-H$ colors place them on the blue edge of
the T/Y sequence in color--magnitude diagrams, and their
integrated-light $J-K$ color of $-1.06\pm0.26$\,mag is normal for
objects of comparable absolute magnitude but would be relatively blue
for a late-T dwarf.

On mid-infrared color--magnitude diagrams, the integrated-light
photometry of \objAB\ follows the sequence of late-T and Y~dwarfs, but
its location is unusual given the integrated-light spectral type of
T9.  \objAB's integrated-light color $[3.6]-[4.5] = 2.42\pm0.07$\,mag
is the reddest of any known T9 dwarf, with the next reddest objects
being the T8.5+Y0 binary \ace{AB} (T9 in integrated-light with
$[3.6]-[4.5] = 2.33\pm0.03$\,mag) followed by \flolong\ (T9) and
\suelong\ (T9p) both with $[3.6]-[4.5] = 2.22\pm0.04$\,mag.  The only
late-T dwarf that is as red as \objAB\ is WISE~J081117.81$-$805141.3
(T9.5:, $[3.6]-[4.5] = 2.42\pm0.07$\,mag), but like other T9 and T9.5
dwarfs it is $\approx$1.0--1.5\,mag brighter in $[3.6]$-band absolute
magnitude than the combined light of \objAB.  The situation is similar
but less extreme in $J-W2$ color versus $W2$-band absolute magnitude.
\objAB\ is the reddest late-T dwarf ($J-W2 = 5.61\pm0.10$\,mag) with
the exceptions of \jeblong\ (T9p; $5.64\pm0.04$\,mag) and \sue\ (T9p;
$5.86\pm0.11$\,mag).  Both of these were typed as peculiar by
\citet{2011ApJS..197...19K} due to excess flux at $Y$ band and less
flux at $K$ band, similar to features seen in the integrated-light
spectrum of \objAB.  However, unlike these T9p objects, the
integrated-light $W2$-band flux of \objAB\ is 0.7--0.8\,mag fainter,
implying that the individual components are $\gtrsim$1\,mag fainter.

Both components of \objAB\ have very faint absolute magnitudes given
their spectral types.  \objA\ is the faintest T9 dwarf known in any
near- or mid-infrared bandpass, and \objB\ is as faint or fainter than
\zoe\ (Y0p) in near-infrared magnitudes and could be the faintest in
the mid-infrared, too, depending on the binary flux ratio at those
wavelengths.  \objA\ is the only late-T dwarf wholly overlapping in
colors \emph{and} magnitudes with the Y~dwarfs.

\subsection{Bolometric Magnitudes \label{sec:mbol}}

\citet{2013Sci...341.1492D} showed that summing fluxes across the
near- and mid-infrared can produce an estimate for the bolometric
magnitudes (\mbol) of late-T and Y~dwarfs that is only weakly
dependent on the assumed model atmospheres.  We applied this
``super-magnitude'' approach to the integrated-light photometry of
\objAB\ and found $\mbol = 21.58\pm0.12$\,mag. To apportion this
between the two individual components we employed our resolved
near-infrared photometry. For our purposes, only the relative
near-infrared super-magnitude between the two components is relevant,
and we find $\Delta{m_{YJH}} = 0.95\pm0.05$\,mag. To convert this into
a bolometric flux ratio requires an estimate of the bolometric
correction for each component.  \citet{2013Sci...341.1492D} report
bolometric corrections of BC$_{YJH} = 1.6\pm0.6$\,mag at T9 and
$0.8\pm0.6$\,mag at Y0, and we adopt these values for the primary and
secondary, respectively. Thus, the bolometric flux ratio is calculated
as $\Delta{m_{\rm bol}} = \Delta{m_{YJH}} + {\rm BC}_{YJH}({\rm Y0}) -
{\rm BC}_{YJH}({\rm T9}) = 0.2\pm0.8$\,mag, and we find bolometric
magnitudes of $\mbol = 22.2\pm0.4$\,mag and $22.4\pm0.4$\,mag for the
primary and secondary, respectively.

We compare our apparent bolometric magnitude estimates to the typical
absolute magnitudes of late-T and Y dwarfs in order to illustrate an
unusual property of \objAB. Using the \Lbol\ values from Table~S5 of
\citet{2013Sci...341.1492D}, we find that $\Mbol = 20.1\pm0.5$\,mag
for T9 and $\Mbol = 21.06\pm0.25$\,mag for Y0, where the error bars
represent the intrinsic rms scatter in bolometric magnitude among
objects of the same spectral type. Assuming spectral types of T9+Y0
for \objAB, the integrated-light absolute bolometric magnitude would
be $\Mbol = 19.7\pm0.4$\,mag if they were typical of other known
objects. However, this is 1.7\,mag different from the bolometric
magnitude of $\Mbol = 21.4^{+0.4}_{-0.3}$\,mag derived from the
photometry above and the parallactic distance of \dpc\ from
\citet{2014ApJ...783...68B}. Therefore \objAB\ is significantly less
luminous than other known objects of the same spectral type, which at
least partly explains its unusual appearance on color--magnitude
diagrams as described above.

We consider the alternative that the parallax of $94\pm14$\,mas from
\citet{2014ApJ...783...68B} is in error.  They used multiple
telescopes (\WISE, \Spitzer, and Keck) and bandpasses ($H$, $[3.6]$,
$[4.5]$) to measure the parallax, which introduces the possibility of
systematic errors in the astrometry.  This is especially true in the
case of a T9+Y0 binary where the flux ratio could vary considerably
from the near- to mid-infrared causing unaccounted for shifts in
center-of-light measurements at different epochs.  However, for the
components of \objAB\ to have normal magnitudes for a T9+Y0 dwarf pair
would require a distance of $26^{+8}_{-6}$\,pc, where the error bars
account for the uncertainty in the \mbol\ and scatter in \Mbol\ as a
function of spectral type \citep{2013Sci...341.1492D}, and thus a
parallax of $\approx$40\,mas.  This would require photocenter shifts
of more than half the binary separation, and such shifts would have to
be correlated with the parallax factor.  Figure~17 of
\citet{2014ApJ...783...68B} shows that there is both Keck and
\Spitzer\ data on both sides of the ellipse, so a large error in the
parallax due to mixing bandpasses seems unlikely.

We also consider whether orbital motion could have impacted their
parallax.  Because proper motion accommodates for any linear orbital
motion, only acceleration is a concern, and this usually shows up in
the residuals of the fit.  \citet{2014ApJ...783...68B} report $\chi^2
= 23.0$ (27 degrees of freedom), so there are no significant residuals
due to orbital motion, but such acceleration could in principle be
aligned with the parallax motion by chance.  To test this possibility,
we fitted just the \citet{2014ApJ...783...68B} astrometry spanning
2012~October to 2013~November, which is contemporaneous with our Keck
data that show only a small amount of orbital motion, $(\dot{\alpha},
\dot{\delta}) = (-4.3\pm3.9, 3.9\pm2.3)$\,mas\,yr$^{-1}$. We find a
relative parallax of $89\pm18$\,mas in this fit, which is consistent
with their reported value.  Finally, we have determined a preliminary
parallax of $130\pm38$\,mas from our own ongoing astrometry program at
the Canada-France-Hawaii Telescope \citep{2012ApJS..201...19D}, which
further supports the unusually low luminosity of \objAB.

\subsection{Estimated Physical Properties \label{sec:prop}}

We used the cloud-free Cond evolutionary models of
\citet{2003A&A...402..701B} to estimate the physical properties of the
binary components from their luminosities and simple assumptions for
the system age. We combine our bolometric magnitudes from above and
the parallactic distance from \citet{2014ApJ...783...68B} to compute
$\log(\Lbol/\Lsun) = -6.95\pm0.20$\,dex for the primary and
$-7.01\pm0.22$\,dex for the secondary. We interpolate the Cond model
grid at ages of 1\,Gyr and 10\,Gyr to derive masses, temperatures, and
other properties from these \Lbol\ values and present the results in
Table~\ref{tbl:prop}.

Some properties that we derive from evolutionary models vary
significantly with the assumed age, while others do not. For example,
the primary component's mass and effective temperature are
$4.6^{+1.0}_{-1.1}$\,\Mjup\ and $320^{+35}_{-40}$\,K at 1\,Gyr but
$16.9^{+3.8}_{-4.0}$\,\Mjup\ and $345\pm45$\,K at 10\,Gyr. This is
because models predict a radius contraction of only 17\% from 1\,Gyr
to 10\,Gyr, so for a given \Lbol\ the \Teff\ estimate changes by a
small amount.  Unlike the radius however, luminosity is a strong
function of mass and age at field ages \citep[$\Lbol \propto M^{2.4}
t^{1.25}$;][]{2001RvMP...73..719B}.  Consequently, a property like
surface gravity that depends on mass changes significantly with the
assumed age ($\approx$0.7\,dex) while the effective temperature only
changes by $\lesssim$25\,K.

\subsubsection{High Surface Gravity \label{sec:logg}}

There is an extensive literature on the impact of both surface gravity
and metallicity variations on the emergent spectra of T dwarfs with
$\Teff = 600$--1000\,K \citep[e.g.,][]{2002ApJ...573..394B,
  2007ApJ...660.1507L, 2007ApJ...667..537L, 2013ApJ...777...36M}, but
similar studies are only beginning for $\lesssim$500\,K objects at the
T/Y transition.  The latest model atmospheres from
\citet{2014ApJ...787...78M} show how the shapes of normalized near-IR
flux peaks change with varying surface gravity (their Figure~15) at
the temperatures of Y~dwarfs.  Models at $\logg = 5.0$\,dex ($g$ in
cgs units) have distinctive features compared to $\logg = 4.0$\,dex at
a common $\Teff = 450$\,K such as substantially increased flux on the
blue side of the peak-normalized $Y$ band, a broader $J$-band peak
caused by increased flux on the red side, and a wavelength for the
$H$-band flux peak slightly shifted to the red. Note that the $Y$-band
excess is only a change in the shape of the peak and not its
amplitude, as \citet{2014ApJ...787...78M} predict that $Y-J$ actually
becomes $\approx$0.5\,mag redder due to the integrated flux at $Y$
band being more suppressed than at $J$ band when going from $\logg =
4.0$\,dex to 5.0\,dex.

We observe all of these features in our integrated-light spectrum of
\objAB\ (Figure~\ref{fig:spectrum}). To our knowledge, this is the
first object known to possess all of these traits.  A number of
objects have been reported to display enhanced flux on the blue side
of $Y$ band \citep[e.g.,][]{2011ApJS..197...19K, 2014ApJ...780...62L},
but of these only \vallong\ (Y0p) has been reported to show the
slightly increased flux on the red side of the $J$-band peak
\citep{2014MNRAS.444.1931P}.  \zoe\ (Y0p) displays the shifted peak
wavelength in $H$ band \citep{2011ApJ...743...50C} but has a normal
$J$ band and no $Y$-band data.  Unfortunately, non-solar metallicity
models are not yet available at these temperatures, so it is not known
if any of these traits could also be reproduced by a sub-solar
metallicity, as is often the case for earlier type T dwarfs where the
effects of low metallicity and high gravity can be similar
\citep[e.g.,][]{2006ApJ...639.1095B, 2013ApJ...777...36M}.

We conclude therefore that \objAB\ is very likely to be relatively
old, having higher surface gravity and/or lower metallicity compared
to other known T/Y transition objects. The chief caveat is that we
only have an integrated-light spectrum that is likely dominated by the
primary along with our resolved broadband photometry to characterize
the photospheres of the individual components. Our spectral
decomposition analysis in Section~\ref{sec:spt} demonstrates that even
a $\approx$1\,mag fainter secondary can influence the best-fit
template match, since in this case it was the Y0p \zoe.  Among the
next tier of best matching templates both components of \ace{AB}
(T8.5+Y0) were often favored, and these are also high surface
gravity/low metallicity candidates \citep{2012ApJ...758...57L,
  2014ApJ...780...62L}.

\subsubsection{Semimajor Axis and Orbital Period \label{sec:sma}}

The projected separation of \objAB\ at discovery was $\rho =
0\farcs0875\pm0\farcs0021$ ($0.93^{+0.12}_{-0.16}$\,AU), and we can
use Table~6 of \citet{me-ecc} to convert this into a statistical
estimate of the semimajor axis ($a$). The conversion factor for
very-low mass visual binaries ($a/\rho = 1.16^{+0.81}_{-0.31}$) gives
a semimajor axis estimate of $0\farcs10^{+0.03}_{-0.04}$
($1.1^{+0.4}_{-0.5}$\,AU). However, because we discovered \objAB\ very
near the resolution limit of our Keck LGS AO images, we are likely in
the ``moderate discovery bias'' case where the inner working angle of
the discovery observations (${\rm IWA} \approx 0\farcs055$) is roughly
half the size of the semimajor axis and a different conversion factor
applies.  Thus, our best estimate for the semimajor axis is
$0\farcs095^{+0.023}_{-0.031}$ ($1.0^{+0.3}_{-0.4}$\,AU).  We note
that \objAB\ likely has the smallest projected separation in AU at
discovery of any visual binary ever found among L, T, or Y dwarfs.
The next tightest is 2MASS~J15344984$-$2952274AB that was discovered
at $1.01\pm0.03$\,AU \citep{2003ApJ...586..512B}.

We can further estimate the orbital period of \objAB\ via Kepler's
Third Law using the total system mass estimate we derived from
evolutionary models. Assuming an age of 10\,Gyr gives $\Mtot =
32^{+5}_{-6}$\,\Mjup\ and using the smaller of the semimajor axes
above (moderate discovery bias) yields a period estimate of $P =
5.9^{+2.0}_{-3.1}$\,yr.  Even if \objAB\ were much younger and thereby
lower mass ($\Mtot = 8.7^{+1.3}_{-1.6}$\,\Mjup\ for an age of 1\,Gyr),
its estimated orbital period would still be quite short
($11^{+4}_{-6}$\,yr).  Thus, regardless of the current uncertainties
in the system mass and semimajor axis, \objAB\ appears to be an
excellent candidate for dynamical mass determination via monitoring of
the binary orbit over the next several years, as it typically requires
only $\approx$30\% coverage of the orbital period to enable a direct
measurement of the system mass \citep[e.g.,][]{2009ApJ...706..328D}.
In fact, we already detect a small amount of orbital motion in our
Keck LGS AO imaging that spans 1.1\,yr ($7\pm3$\,\masyr).

\section{Discussion}

Both components of \objAB\ appear to be less luminous than any other
known members of substellar binaries, even the Y~dwarf companions
\eli{B} and \ace{B} \citep{2011ApJ...740..108L, 2012ApJ...758...57L}.
This in turn implies that many of the properties of \objAB\ are
extreme relative to other known systems. To reach such low
luminosities, both components of \objAB\ must be planetary mass
($\lesssim$13\,\Mjup) and/or very old. We identify several
spectrophotometric features that are indicative of high surface
gravity, so it appears likely that the system is indeed old, although
sub-solar metallicity models at $\Teff = 300$--500\,K are needed to
strengthen this conclusion.  We estimate component masses that are
quite small, with Cond models giving 13--21\,\Mjup\ for the primary
(1$\sigma$ range) and 12--19\,\Mjup\ for the secondary even if the
system is 10\,Gyr old.  Currently, a large source of uncertainty in
the component properties is how the integrated-light bolometric flux
should be divided between the two components. Resolved mid-infrared
photometry and near-infrared spectroscopy are needed to more precisely
determine the physical properties of both components.

The faintness of \objA\ is unexpected given its T9 spectral type.  Its
absolute magnitudes are $\approx$1.0--1.5\,mag fainter in the
mid-infrared and $\approx$2\,mag fainter in the near-infrared than is
typical for its spectral type, even allowing for a $\pm$0.5 subtype
uncertainty in classification.  The faint absolute magnitudes lead to
a low bolometric flux estimate, and in turn a low temperature of
$\Teff = 345\pm45$\,K for the primary (assuming an age of 10\,Gyr).
This is $\approx$100\,K colder than any other known late-T
dwarf.\footnote{The lowest published \Lbol-based \Teff\ for a late-T
  dwarf is $502\pm10$\,K for \joe\ (T9), assuming an age of 1\,Gyr
  \citep{2013Sci...341.1492D}.  Examining other objects with recently
  published parallaxes, we compute that only \jeb\ (T9p), \sue\ (T9p),
  and \gus\ (T9.5) have cooler \Teff\ estimates, ranging from
  $461\pm13$\,K to $466\pm13$\,K, assuming ages of 1\,Gyr and using
  parallaxes from \citet{2014ApJ...796...39T} and photometry from
  \citet{2011ApJS..197...19K}.}  According to the Stefan-Boltzmann
Law, at a given luminosity $\Teff \propto R^{1/2}$.  Thus, a
$\approx$20\% discrepancy in \Teff\ could be compensated by the radius
being 40\% smaller, i.e., 0.55\,\Rjup\ instead of 0.91\,\Rjup.  Such a
radius is unphysical as it is significantly smaller than any predicted
or measured substellar radii \citep[e.g.,][and references
therein]{2009AIPC.1094..102C, 2014arXiv1411.4047M}.  According to
evolutionary models from \citet{2008ApJ...689.1327S}, assuming an age
older than 10\,Gyr would have only a small effect ($\lesssim$1\%) on
the radius, and assuming a 2$\times$ larger mass would reduce the
radius by only $\approx$10\%.  However, these effects might partially
explain the apparent temperature discrepancy between \objA\ and other
T9/T9.5 dwarfs.  A systematic error in the \citet{2014ApJ...783...68B}
parallax distance could also mitigate this discrepancy, but to explain
it entirely would require an unlikely error of a factor of $\gtrsim$2
in the measured parallax.  Therefore, we conclude that the discrepancy
between \objA's spectral type and its temperature cannot be completely
explained by systematic errors in the assumed radius or luminosity.

One intriguing explanation for the $\approx$100\,K cooler \Teff\ of
\objA\ compared to other objects of similar spectral type is that the
features in the $J$ and $H$ bands that are widely used to determine
spectral types for late-T and Y dwarfs are highly dependent on gravity
or metallicity. Since these flux peaks are shaped by molecular
features (e.g., CH$_4$, H$_2$O, H$_2$, and NH$_3$), a sensitivity to
gravity or metallicity is not unexpected.  This did not appear to be
the case in the initial sample of late-T and Y dwarfs with \Lbol-based
temperatures from \citet{2013Sci...341.1492D} where spectral types
tended to track well with \Teff, with the possible exception of T9.5
dwarfs that seemed to be warmer or similar temperature compared to T9
dwarfs.  However, as the sample of parallaxes for late-T and Y dwarfs
grows, a jumbling of objects with the same \Teff\ but very different
spectral types should become clearly apparent if the range of surface
gravities and/or metallicities of the field population significantly
affect near-infrared spectral types.

With only a few of the coldest known brown dwarfs showing signatures
of high surface gravity, it seems that objects like the components of
\objAB\ must either be relatively uncommon or observationally selected
against. \citet{2014ApJ...787...78M} predict that high surface gravity
objects at a given temperature are indeed $\approx$0.2--0.3\,mag
fainter in the $W2$ band that was used to discover most known objects
of this type. However, this would lead to lower gravities being over
represented by only $\approx$30\%--50\% in a magnitude-limited sample.
In fact, given the generic assumption of a declining star formation
history in the solar neighborhood
\citep[e.g.,][]{2009MNRAS.397.1286A}, older, higher gravity brown
dwarfs should be much more common at a given \Teff\ than lower gravity
ones, unless the mass function compensates by having a strong
preference for producing lower mass brown dwarfs.  If higher gravity
brown dwarfs are uncommon it would imply either an unusual age
distribution for brown dwarfs in the solar neighborhood (favoring
younger ages) or a large undiscovered population of planetary-mass
brown dwarfs that are the older and colder but similar mass
counterparts to the current Y~dwarf sample.

Fortunately, \objAB\ will soon provide an opportunity to much better
constrain its physical properties. With a projected separation of
$0.93^{+0.12}_{-0.16}$\,AU and estimated orbital period
$\lesssim$10\,yr, \objAB\ is not only a contender for the tightest
known ultracool binary, it is also the first binary discovery at the
T/Y transition that is likely to yield a dynamical mass quickly, as
other discoveries have estimated orbital periods of $\sim$40--400\,yr.
The relative orbit of \objAB\ could be determined by astrometric
monitoring within just a few years.  A dynamical mass for \objAB\ will
test the hypothesis that the components are old and massive with high
surface gravity, which is solely based on current theoretical
predictions of the spectrophotometric properties of brown dwarfs. As
perhaps the only planetary-mass brown dwarfs with a dynamical mass
measurement for the next several years, \objAB\ will be a prime target
for \textsl{James Webb Space Telescope} spectroscopy to test cold
model atmospheres where the physical parameters are well constrained
by observations.


\acknowledgments

We are grateful to 
the anonymous referee for a prompt and productive review;
Michael Cushing for providing published spectra of late-T and
Y~dwarfs;
Randy Campbell, Gary Punawai, Terry Stickel, Hien Tran, and the Keck
Observatory staff for assistance with the Keck LGS AO observing;
the Gemini Observatory staff for obtaining the NIRI photometry and
GNIRS spectroscopy through queue observing;
and James R.\ A.\ Davenport for distributing his IDL implementation of
the cubehelix color scheme.
This work was supported by a NASA Keck PI Data Award, administered by
the NASA Exoplanet Science Institute. 
M.C.L.\ acknowledges support from NSF grant AST09-09222.
Our research has employed NASA's Astrophysical Data System and the
SIMBAD database operated at CDS, Strasbourg, France.
Finally, the authors wish to recognize and acknowledge the very
significant cultural role and reverence that the summit of Mauna Kea has
always had within the indigenous Hawaiian community.  We are most
fortunate to have the opportunity to conduct observations from this
mountain.

{\it Facilities:} \facility{Keck:II (LGS AO, NIRC2)} \facility{Gemini:Gillett}

\clearpage


\clearpage

\clearpage
\begin{landscape}
\begin{deluxetable}{lcccccc}
\tablecaption{Keck LGS AO Observations of \objAB \label{tbl:keck}}
\tablewidth{0pt}
\tablehead{
\colhead{Date} &
\colhead{Airmass} &
\colhead{Filter} &
\colhead{FWHM} &
\colhead{$\rho$} &
\colhead{P.A.} &
\colhead{$\Delta{m}$} \\
\colhead{(UT)} &
\colhead{} &
\colhead{} &
\colhead{(mas)} &
\colhead{(mas)} &
\colhead{(\degree)} &
\colhead{(mag)} }
\startdata

 2012 Sep  7 & 1.084 &  $Y$  &     $48\pm26$     & $85.9\pm4.4$ & $265.0\pm4.4$ \phn& $ 0.83\pm0.23 $ \\
 2012 Sep  7 & 1.118 &  $J$  &     $58\pm14$     & $88.2\pm3.4$ & $258.8\pm2.0$ \phn& $ 0.98\pm0.03 $ \\
 2012 Sep  7 & 1.091 &  $H$  &     $52\pm 6$\phn & $87.9\pm3.5$ & $259.5\pm1.8$ \phn& $ 0.97\pm0.10 $ \\
 2012 Sep  7 &\nodata& mean  &      \nodata      & $87.5\pm2.1$ & $259.7\pm1.3$ \phn&    \nodata      \\
             &       &       &                   &              &                   &                 \\
 2012 Oct  8 & 1.093 &  $Y$  &     $64\pm17$     & $91.0\pm2.9$ & $260.5\pm4.1$ \phn& $ 0.83\pm0.23 $ \\
 2012 Oct  8 & 1.236 &  $J$  &     $60\pm16$     & $93.9\pm4.2$ & $262  \pm6  $ \phn& $ 1.07\pm0.08 $ \\
 2012 Oct  8 & 1.198 &  $H$  &     $49\pm 4$\phn & $85.6\pm3.6$ & $258.3\pm4.1$ \phn& $ 1.15\pm0.14 $ \\
 2012 Oct  8 & 1.129 &$CH_4s$&     $55\pm11$     & $93.5\pm4.4$ & $262.0\pm1.2$ \phn& $ 1.32\pm0.13 $ \\
 2012 Oct  8 &\nodata& mean  &      \nodata      & $90.6\pm1.8$ & $261.6\pm1.1$ \phn&    \nodata      \\
             &       &       &                   &              &                   &                 \\ 
 2013 Oct 22 & 1.098 &  $J$  &     $56\pm10$     & $92.9\pm4.0$ & $263.6\pm1.3$ \phn& $ 0.90\pm0.16 $ \\

\enddata

\tablecomments{All photometry on the MKO photometric system.  The
  identical $Y$-band flux ratios are not a typo.}

\end{deluxetable}
\end{landscape}

\begin{deluxetable}{lcccc}
\tablecaption{Gemini-North NIRI Photometry of \obj \label{tbl:gem}}
\tablewidth{0pt}
\tablehead{
\colhead{Filter} &
\colhead{$t_{\rm exp}$} &
\colhead{Photometry} &
\colhead{Date}  \\
\colhead{} &
\colhead{(s)} &
\colhead{(mag)} &
\colhead{(UT)} }
\startdata
  $Y$  & $6\times60$ & $21.60\pm0.15$ & 2013 Jan 10 \\
  $J$  & $6\times60$ & $20.69\pm0.07$ & 2013 Jan 10 \\
  $H$  &$28\times30$ & $21.30\pm0.12$ & 2013 Jan 10 \\
$CH_4s$&$28\times30$ & $20.51\pm0.14$ & 2013 Jan 12 \\
  $K$  &$28\times30$ & $ 22.4\pm0.4 $\tablenotemark{a} & 2013 Jan 10 \\
\enddata

\tablenotetext{a}{Given the very low signal-to-noise ratio at $K$
  band, we suspect our photometric errors in this filter may be
  underestimated.}

\end{deluxetable}

\begin{deluxetable}{lcccccccc}
\setlength{\tabcolsep}{0.040in}
\tabletypesize{\footnotesize}
\tablecaption{Median-Flux Spectral Indices for \objAB\ \label{tbl:indices}}
\tablewidth{0pt}
\tablehead{
\colhead{Spectral} &
\colhead{\objAB} &
\colhead{} &
\multicolumn{4}{c}{Average Median-Flux Values from \citet{2013ApJS..205....6M}} &
\colhead{} &
\colhead{Best-match}  \\
\cline{4-7}
\colhead{Index} &
\colhead{} &
\colhead{} &
\colhead{T8.5} &
\colhead{T9} &
\colhead{T9.5} &
\colhead{Y0} &
\colhead{} &
\colhead{Spec.\ Type} }
\startdata

W$_J$       &  $0.188\pm0.031$  & &  $0.295\pm0.047$  &  $0.203\pm0.038$                   &   $0.145\pm0.021$  &  $0.117\pm0.043$                   & &   T9    \\
$J$-narrow  &  $0.923\pm0.036$  & &  $0.884\pm0.044$  &  $0.879\pm0.053$                   &   $0.865\pm0.071$  &  $0.778\pm0.050$                   & &   T9    \\
CH$_4-J$    &  $0.063\pm0.024$  & &  $0.188\pm0.056$  &  $0.117\pm0.030$                   &   $0.071\pm0.032$  &  $0.045\pm0.031$                   & &   T9.5  \\
CH$_4-H$    &  $0.120\pm0.029$  & &  $0.122\pm0.035$  &  $0.068\pm0.067$                   &$-0.021\pm0.116$\phs&  $0.066\pm0.133$                   & &   T8.5  \\
NH$_3-H$    &  $0.457\pm0.055$  & &  $0.610\pm0.059$  &  $0.539\pm0.054$                   &   $0.443\pm0.101$  &  $0.385\pm0.105$                   & &   T9.5  \\
$Y/J$       &  $0.571\pm0.038$  & &  $0.432\pm0.082$  &  $0.448\pm0.095$                   &   $0.357\pm0.008$  &  $0.423\pm0.131$                   & & \nodata \\
$H/J$       &  $0.539\pm0.027$  & &  $0.503\pm0.045$  &  $0.555\pm0.030$                   &   $0.504\pm0.080$  &  $0.467\pm0.061$                   & & \nodata \\
$K/J$       &  $0.066\pm0.007$  & &  $0.116\pm0.039$  &  $0.100\pm0.032$                   &   $0.033\pm0.083$  &  $0.063\pm0.031$                   & & \nodata \\
$J$-wing    &  $0.322\pm0.043$  & &      \nodata      &  $0.320\pm0.005$\tablenotemark{a}  &        \nodata     &  $0.164\pm0.012$\tablenotemark{a}  & &   T9    \\

\enddata

\tablecomments{All spectral indices except $J$-wing are described in
  \citet{2013ApJS..205....6M}.  We exclude the indices H$_2{\rm O}-J$
  and H$_2{\rm O}-H$ as they are both saturated, i.e., have zero flux
  in the numerator, for \objAB.}

\tablenotetext{a}{These are the values for the spectral standards
  \joe\ (T9) and \bob\ (Y0) reported by
  \citet{2014MNRAS.444.1931P} who originally defined this index.}

\end{deluxetable}

\begin{deluxetable}{lcc}
\tablecaption{Integrated-Light Properties of \objAB \label{tbl:prop1}}
\tablewidth{0pt}
\tablehead{
\colhead{Property} &
\colhead{A+B} &
\colhead{Ref.} }
\startdata
\multicolumn{3}{c}{Measured} \\
\cline{1-3}
$z_{\rm SDSS}$ (mag)                             &   $24.10\pm0.13$\phn        & Lod13                              \\
$Y_{\rm MKO}$ (mag)                              &   $21.60\pm0.15$\phn        & \S\ref{sec:niri}                   \\
$J_{\rm MKO}$ (mag)                              &   $20.69\pm0.07$\phn        & \S\ref{sec:niri}                   \\
$H_{\rm MKO}$ (mag)                              &   $21.30\pm0.12$\phn        & \S\ref{sec:niri}                   \\
$CH_4s_{\rm MKO}$ (mag)                          &   $20.51\pm0.14$\phn        & \S\ref{sec:niri}                   \\
$K_{\rm MKO}$ (mag)                              &   $21.75\pm0.25$\phn        & \S\ref{sec:gnirs}                  \\
$[3.6]_{\rm IRAC}$ (mag)                         &   $17.50\pm0.07$\phn        & Kir12                              \\
$[4.5]_{\rm IRAC}$ (mag)                         &   $15.08\pm0.02$\phn        & Kir12                              \\
Spectral type                                    &          T9p                & \S\ref{sec:gnirs}, \S\ref{sec:spt} \\
Distance (pc)                                    & $10.6^{+1.3}_{-1.8\phn}$    & Bei14                              \\
$\mu_{\alpha}\cos{\delta}$ (\arcsec\,yr$^{-1}$)  &  $-0.441\pm0.013$\phs       & Bei14                              \\
$\mu_{\delta}$             (\arcsec\,yr$^{-1}$)  &  $-0.026\pm0.016$\phs       & Bei14                              \\
\cline{1-3}
\multicolumn{3}{c}{Estimated} \\
\cline{1-3}
\mbol\ (mag)                                     &   $21.58\pm0.12$\phn        & \S\ref{sec:mbol}                   \\
$\log(\Lbol/\Lsun)$                              & $-6.67^{+0.12}_{-0.15}$\phs & \S\ref{sec:mbol}, Bei14            \\
\enddata

\tablerefs{References: \S\ numbers refer to this paper; Bei14
  \citep{2014ApJ...783...68B}; Kir12 \citep{2012ApJ...753..156K};
  Lod13 \citep{2013A&A...550L...2L}.}

\end{deluxetable}

\begin{deluxetable}{lcc}
\tabletypesize{\small}
\tablecaption{Resolved Properties of \objAB \label{tbl:prop}}
\tablewidth{0pt}
\tablehead{
\colhead{Property} &
\colhead{Component A} &
\colhead{Component B} }
\startdata

\multicolumn{3}{c}{Measured} \\
\cline{1-3}
Separation\tablenotemark{a}     & \multicolumn{2}{c}{$0\farcs0875\pm0\farcs0021$              } \\
P.A.\tablenotemark{a}           & \multicolumn{2}{c}{   $259\fdg7\pm1\fdg3$\phn\phn           } \\
$\Delta{Y_{\rm MKO}}$ (mag)     & \multicolumn{2}{c}{       $0.83\pm0.16$\tablenotemark{b}    } \\
$\Delta{J_{\rm MKO}}$ (mag)     & \multicolumn{2}{c}{       $0.99\pm0.03$\tablenotemark{b}    } \\
$\Delta{H_{\rm MKO}}$ (mag)     & \multicolumn{2}{c}{       $1.03\pm0.08$\tablenotemark{b}    } \\
$\Delta{CH_4s_{\rm MKO}}$ (mag) & \multicolumn{2}{c}{       $1.32\pm0.13$                     } \\
$Y_{\rm MKO}$ (mag)             &    $22.02\pm0.16$             &    $22.85\pm0.19$             \\
$J_{\rm MKO}$ (mag)             &    $21.06\pm0.07$             &    $22.05\pm0.07$             \\
$H_{\rm MKO}$ (mag)             &    $21.66\pm0.12$             &    $22.69\pm0.14$             \\
$CH_4s_{\rm MKO}$ (mag)         &    $20.79\pm0.14$             &    $22.11\pm0.17$             \\
Spectral type                   &           T9                  &           Y0                  \\

\cline{1-3}
\multicolumn{3}{c}{Estimated} \\
\cline{1-3}
$\Delta\mbol$ (mag)             & \multicolumn{2}{c}{        $0.2\pm0.8$                      } \\
\mbol\ (mag)                    &     $22.2\pm0.4$\phn          &     $22.4\pm0.4$\phn          \\
$\log(\Lbol/\Lsun)$             &    $-6.95\pm0.20$\phs         &    $-7.01\pm0.22$\phs         \\

\cline{1-3}
\multicolumn{3}{c}{Model-derived (Cond, $t = 1$\,Gyr)} \\
\cline{1-3}
\Mtot\ (\Mjup)                  & \multicolumn{2}{c}{         $8.7^{+1.3}_{-1.6}$             } \\
Mass (\Mjup)                    & $   4.6^{+1.0}_{-1.1}$        &    $4.3^{+1.0}_{-1.2}$        \\
\Teff\ (K)                      & $    320^{+35}_{-40}$         & $   310^{+35}_{-40}$          \\
Radius (\Rjup)                  & $ 1.0674^{+0.0027}_{-0.0017}$ & $ 1.0675^{+0.0029}_{-0.0018}$ \\
\logg\ (cgs)                    & $   3.99^{+0.11}_{-0.09}$     & $  3.96^{+0.12}_{-0.11}$      \\

\cline{1-3}
\multicolumn{3}{c}{Model-derived (Cond, $t = 10$\,Gyr)} \\
\cline{1-3}
\Mtot\ (\Mjup)                  & \multicolumn{2}{c}{      $32^{+5}_{-6}$                     } \\
Mass (\Mjup)                    & $  16.9^{+3.8}_{-4.0}$        & $   15.9^{+3.5}_{-4.4}$       \\
\Teff\ (K)                      & $   345\pm45$                 & $   330\pm45$                 \\
Radius (\Rjup)                  & $ 0.913^{+0.023}_{-0.025}$    & $ 0.919^{+0.034}_{-0.015}$    \\
\logg\ (cgs)                    & $  4.69^{+0.13}_{-0.11}$      & $  4.65^{+0.14}_{-0.12}$      \\

\enddata

\tablenotetext{a}{Weighted average of measurements in different
  filters at epoch 2012~Sep~7~UT.}

\tablenotetext{b}{Weighted average of measurements from multiple
  epochs.}

\end{deluxetable}

\end{document}